\documentclass[sigconf,screen]{acmart}
\AtBeginDocument{%
  \providecommand\BibTeX{{%
    \normalfont B\kern-0.5em{\scshape i\kern-0.25em b}\kern-0.8em\TeX}}}

\setcopyright{none}
\copyrightyear{2024}
\acmYear{2024}
\acmDOI{XXXXXXX.XXXXXXX}

\acmConference[ToMinHAI at CHI 2024]{Workshop on Theory of Mind in Human-AI Interaction at CHI 2024}{May 12th}{Honolulu, Hawaii}
\begin{document}

\title{Mutual Theory of Mind for Human-AI Communication}

\author{Qiaosi Wang}
\email{qswang@gatech.edu}
\affiliation{%
  \institution{Georgia Institute of Technology}
  \city{Atlanta}
  \state{GA}
  \country{USA}
}

\author{Ashok K. Goel}
\email{ashok.goel@cc.gatech.edu}
\affiliation{%
  \institution{Georgia Institute of Technology}
  \city{Atlanta}
  \state{GA}
  \country{USA}
}

\renewcommand{\shortauthors}{Wang and Goel}

\begin{abstract}
New developments are enabling AI systems to perceive, recognize, and respond with social cues based on inferences made from humans' explicit or implicit behavioral and verbal cues. These AI systems, equipped with an equivalent of human's Theory of Mind (ToM) capability, are currently serving as matchmakers on dating platforms, assisting student learning as teaching assistants, and enhancing productivity as work partners. They mark a new era in human-AI interaction (HAI) that diverges from traditional human-computer interaction (HCI), where computers are commonly seen as tools instead of social actors. Designing and understanding the human perceptions and experiences in this emerging HAI era becomes an urgent and critical issue for AI systems to fulfill human needs and mitigate risks across social contexts. In this paper, we posit the Mutual Theory of Mind (MToM) framework, inspired by our capability of ToM in human-human communications, to guide this new generation of HAI research by highlighting the iterative and mutual shaping nature of human-AI communication. We discuss the motivation of the MToM framework and its three key components that iteratively shape the human-AI communication in three stages. We then describe two empirical studies inspired by the MToM framework to demonstrate the power of MToM in guiding the design and understanding of human-AI communication. Finally, we discuss future research opportunities in human-AI interaction through the lens of MToM. 

\end{abstract}

\ccsdesc[500]{Human-centered computing~HCI theory, concepts and models}
\ccsdesc[500]{Human-centered computing~Empirical studies in HCI}
\ccsdesc[300]{Human-centered computing~Natural language interfaces}
\ccsdesc[300]{Computing methodologies~Artificial intelligence}

\keywords{theory of mind, human-AI interaction, social intelligence}



\maketitle

\section{Introduction}
With new technology advancements, AI systems are increasingly serving different social roles across contexts. For example, AI systems are acting as matchmakers to provide matches for our business or life partners, as personal assistants to manage our daily routines, as learning assistants to facilitate student learning, and more. These AI systems are often able to perceive, recognize, and react to human characteristics, needs, and perceptions embedded in our behavioral and verbal cues. This presents a new interaction paradigm in Human-AI Interaction (HAI) that diverges from the traditional Human-Computer Interaction (HCI)--- people are expecting AI systems to possess social intelligence for diverse social functions, yet are often uncertain about the AI's capabilities and social roles during interactions. Oftentimes during HAI, people build different perceptions or mental models of the AI based on the AI outputs~\cite{Gero2020MentalSetting,weisz2021perfection,wang2019human,wang2021towards}; at the same time, AI systems are also constantly building different interpretations of human characteristics, needs, and goals based on human's input~\cite{Pynadath2005PsychSim:Agents,si2010modeling,divekar2019you,walsh2022understanding,eicher2017toward}. In this emerging HAI paradigm, interpretations of each other, from both the human side and the AI side, are playing an increasingly crucial part in shaping human-AI interactions, but how should the HAI community study it in a systematic way to enhance human-AI interactions?

In this paper, we propose viewing this emerging HAI paradigm through the lens of Mutual Theory of Mind (MToM). We draw inspirations from the human-human communication process, largely enabled by our basic cognitive and social ability of Theory of Mind (ToM). ToM is our ability to make conjectures about ourselves and others' mental states (e.g., emotions, intentions)~\cite{Baron-cohen1999EvolutionMind,gopnik1992child}, and it is the key to enabling many human social behaviors such as communication repair and making shared plans or goals~\cite{baron1985does}. Having the capability of ToM enables us to construe a mental model of others' minds, which includes their thoughts, preferences, goals, needs, plans, etc.~\cite{Baron-cohen1999EvolutionMind,premack1978does}. In typical human-human communications, having a \textbf{Mutual Theory of Mind (MToM), meaning all parties involved in the interaction possess the ToM}, enables us to continuously refine our interpretations of each others' minds through behavioral and verbal feedback, helping us to maintain constructive and coherent communications.

Drawing on the parallel between the MToM in human-human communications and the emerging HAI paradigm where both humans and AIs can construct representations of each other during communications, we propose the framework of Mutual Theory of Mind to guide the next generation of research in human-AI communications. We argue that~\textit{MToM as a framework provides a process and content account of human-AI communication} that emphasizes the iterative mutual shaping of each party's interpretations and feedback through different stages of the communication process. We will first review relevant literature on ToM and human-AI communication, then describe the MToM framework in details. We then summarize two of our empirical studies inspired by the MToM framework, focusing specifically on the second-level ToM--- the idea of ``I can think about what you think about my mind''--- in human-AI communications. Finally, we discuss potential research opportunities in human-AI interaction through the lens of MToM.

\section{Related Work}
\subsection{Theoretical Perspectives of Communication}
Communication is commonly defined as~\textit{``the process of transmitting information and common understanding from one person to another.''}~\cite{lunenburg2010communication} Scholars across disciplines have offered different perspectives to study and enhance communication. 

In communication studies, researchers have focused on the different components at play during the communication process. The classic Shannon-Weaver model of communication~\cite{shannon1948mathematical} outlines several key components during the communication process~\cite{lunenburg2010communication}:~\textit{sender} who initiates the communication process by sending messages~\textit{encoded} using symbols, gestures, words, or sentences through a chosen~\textit{channel} to the~\textit{receiver}. While the message is transmitting through the channel, there could be~\textit{noises} that could distort the message. After receiving the message from the sender, the receiver will~\textit{decode} the message into meaningful information, depending on how the receiver interprets the message. Finally, the receiver will provide~\textit{feedback} as a response to the sender. These key components determine the quality and effectiveness of the communication. 

The Cognitive Science perspective of communication highlights the critical role of ToM~\cite{premack1978does}. ToM enables us to make suppositions of other's minds through verbal and behavioral cues, acting as the foundation of human-human communication~\cite{Baron-cohen1999EvolutionMind,baron1985does}. From this perspective, both interlocutors during communication can form interpretations of what's on the other interlocutor's mind based on the implicit and explicit communication cues. For example, we can often infer the interlocutors' goals, plans, or preferences based on what they said, their facial expressions, or their bodily expressions~\cite{premack1978does,Baron-cohen1999EvolutionMind}. Based on that interpretation we formed about the other's mind, we will act accordingly to correct, explain, or persuade. This cycle of building an interpretation of other's minds and then act upon that interpretation continues iteratively throughout the communication process. Inferring about each other's minds through behavioral cues, according to this perspective, is therefore crucial to a smooth and successful communication. 

Communication process can also be interpreted from the social science perspective through impression management~\cite{Goffman1978TheLife}. In his seminal work,~\citeauthor{Goffman1978TheLife} describes social interaction as an information game between individuals and their audience to maintain the ``veneer of consensus'' to keep the conversation going and to avoid awkwardness. During social interactions, the audience usually try to gather as much information as they could about the individuals they interact with in order to elicit a desirable response from the individual; whereas individuals put up performances through two kinds of expressions—-- expressions that are intentionally performed to leave a certain impression (expression given) or expressions that are unintentionally given off that could influence the audience’s impressions of them (expression given off)—-- to manage impressions~\cite{Goffman1978TheLife}. Throughout interactions, each party conveys their definition of the situation through communications: individuals by expressions and audience by reactions to the individuals. 

These three perspectives on communication emphasize different aspects of the communication process: the communication study perspective focuses on the encoding and decoding process of messages; the cognitive science perspective discusses how behavioral cues can inform our interpretations of interlocutor's minds; the social science perspective describes how interpretations of others' minds could predict our behaviors. Our Mutual Theory of Mind framework attempts to bring these different emphasis together into one coherent framework to understand the mutual shaping process of interpretations and feedback during communication.

\subsection{Theory of Mind in Human-AI Communication}
Over the years, many researchers have recognized the crucial role of ToM in HAI. In human-robot teaming research, ToM has been intentionally built in as part of the system architecture to help robots monitor world state as well as the human state~\cite{Devin2016AnExecution}, to construct simulation of hypothetical cognitive models of the human partner to account for human behaviors that deviate from original plans~\cite{Pynadath2005PsychSim:Agents}, and to help robots to build mental models about user beliefs, plans and goals~\cite{Kim2009TowardsMind,Harbers2009ModelingMind}. Robots built with ToM have demonstrated positive outcomes in team operations~\cite{Devin2016AnExecution} and are perceived to be more natural and intelligent~\cite{Lin2010ReflexivelyAttention}.

Other research in HCI and human-centered AI has also been exploring along the realm of ToM, focusing mostly on enhancing user's mental models and understanding of the AI systems. Prior research has explored people's mental model of AI systems--- people's mental model of AI agents could include global behavior, knowledge distribution, and local behavior~\cite{Gero2020MentalSetting}. People's perception of AI systems is instrumental in guiding how they interact with AI systems~\cite{Gero2020MentalSetting} and thus serves as a precursor to their expectation of AI's behavior. Some recent research has also begun to examine how to automatically infer user's mental model of AI. Prior research suggests the potential of leveraging linguistic cues to indicate people's perception of AIs during human-AI interactions. Researchers have been able to infer users' emotions towards an AI agent~\cite{skowron2011good} and signs of conversation breakdowns~\cite{Liao2018AllWild}from communication cues.

Given that AI's behavior and output could also influence user's mental model of the AI, and therefore how the user decides to interact with the AI, we want to highlight that the interpretation-feedback loop is mutual during the human-AI communication process--- user's mental model of the AI can be informed by the AI's output, yet AI's interpretation of the user can also be informed by the user's output, which is determined by the user's mental model of the AI. We propose the Mutual Theory of Mind framework to capture this mutual shaping process of interpretation-feedback during human-AI communication. 

\section{Mutual Theory of Mind Framework}
\begin{figure*}[t]
  \centering
  \includegraphics[width=0.6\linewidth]{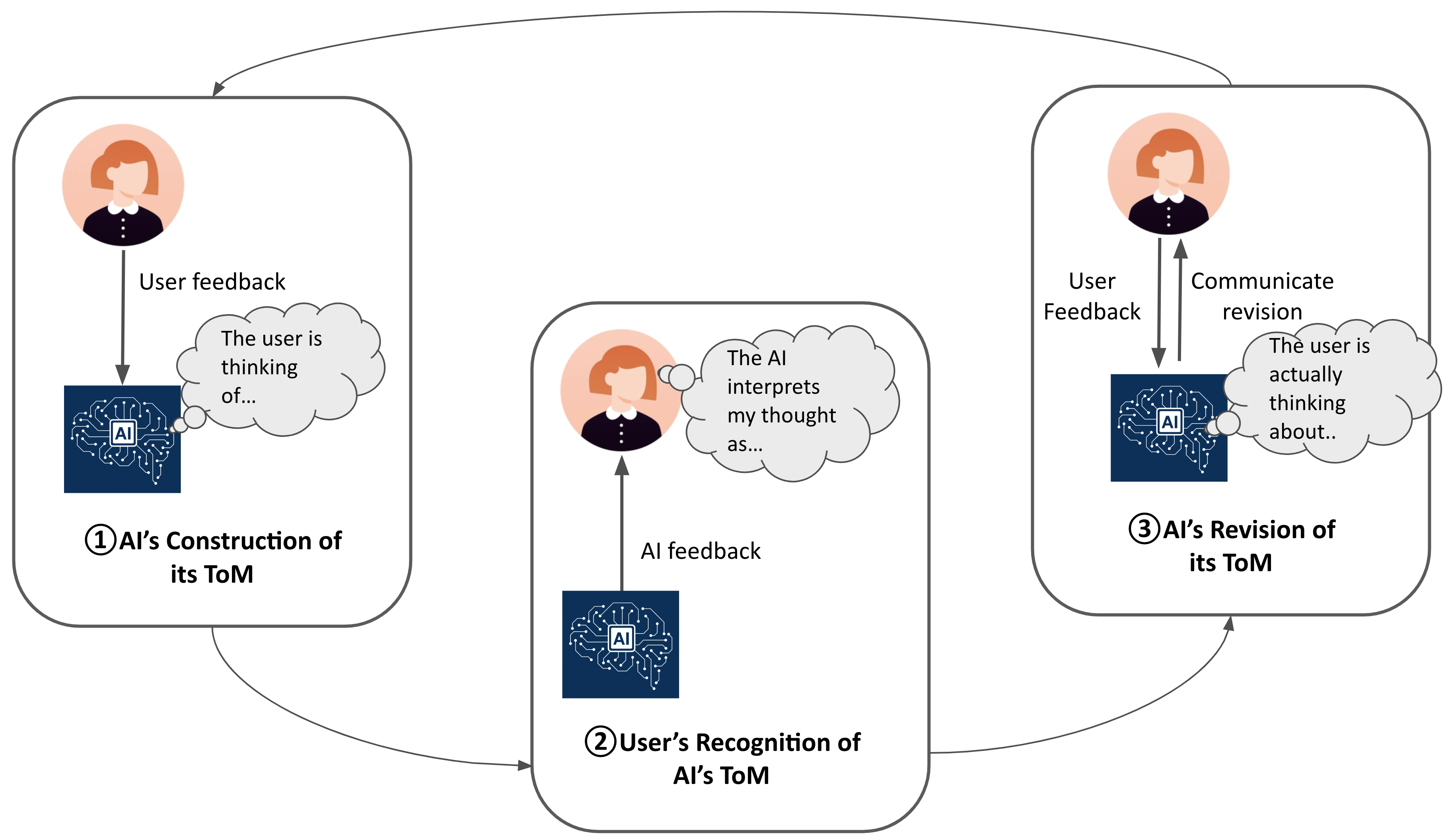}
  \caption{The Mutual Theory of Mind framework that illustrates the iterative process of human-AI communication. The figure also illustrates the three elements across the three stages in MToM.} ~\label{fig:mtom-framework}
\end{figure*}

Drawing from theoretical and empirical work, we posit the MToM framework to guide the understanding and design of communications between humans and AI systems that exhibit social behaviors enabled by ToM-like capability. The MToM framework provides both process and content account of human-AI communication by highlighting~\textit{three elements} that mutually shape the human-AI communication process in~\textit{three stages}. 

\subsection{Three Elements of the MToM Framework}
In the MToM framework, three elements are critical for humans and AI to reach mutual understanding during the communication process:~\textit{interpretation, feedback, and mutuality}. 

In human-AI communication, humans and AI can each construct and revise their~\textit{interpretations} of each other based on feedback from the other party. These interpretations are their interpretations of what's on the other party's mind. This can include the other party's understanding or representations of the world, of the task, of the interlocutor, and of other things. For example, humans can build an interpretation of the AI's representations of the world, of the human, of the plans and goals of the current task during human-AI collaboration, and vice versa. It is important to note that these interpretations can be recursive given that ToM can be of higher levels, meaning that interpretations can not only refer to ''my interpretation of your mind'', but can also refer to ``my interpretation of your interpretation of my mind.'' We further illustrate this recursive property of MToM through our empirical studies in Section~\ref{empirical_mtom}, both focused on second-level ToM. 

\textit{Feedback}, often in the form of verbal or behavioral cues, is generated with different complexities based on the interpretations of each other. For instance, in a human-chatbot conversation, humans would generate simpler command when they believed the chatbot could not understand complex human language; the chatbot would generate simpler feedback when they interpreted the human needs were simple (e.g., asking about the weather). 

While each party involved in the communication is capable of constructing interpretations and generating feedback on their own, communication is a two-way interaction, which means all parties involved in the communication process are~\textit{mutually shaping} each other's interpretations of each other's minds through feedback. Human's interpretations of the AI is constantly shaped by the AI's output, which is shaped by the AI's interpretations of the human's mind, and vice versa. 

These three key elements play a critical part in determining the success and failure of a communication--- inaccurate feedback could inform inaccurate interpretations, inaccurate interpretations could generate inaccurate feedback. Failure in any of them can undermine the mutual shaping process in human-AI communication.

\subsection{Three Stages of MToM Framework}
In the MToM framework, these three elements are constantly shaping the communication between humans and AI during three stages: \textit{AI's construction of its ToM, user's recognition of AI's ToM, and AI's revision of its ToM.} 

In the first stage,~\textit{AI's construction of its ToM}, the AI system takes in user feedback and tries to interpret what's on the user's mind. Depending on the specific communication context, this could be the user's goals, needs, preferences etc. Based on the interpretation, the AI~\textit{constructs} its theory of the user's mind, which helps the AI to generate responses accordingly to help the user fulfill their goals and needs in this instance of communication. 

After the AI generates its response to the user, the user then ~\textit{recognizes} the AI's interpretation of the user's mind based on AI's response. This recognition leads the user to construct their interpretation of the AI, which includes the AI's capability,  working mechanism, and how they are interpreted by the AI. 

The AI's interpretation of the user's mind might not always be accurate. Based on the user's interpretation of the AI, the user provides feedback to the AI, which the AI takes in to ~\textit{revise} or update its interpretation of the user's mind based on the user's feedback. To make sure that the user's interpretations of the AI is accurate after such revisions, it is crucial for the AI to also communicate this revision of its ToM back to the user through feedback. 

Throughout these three stages, the three elements of MToM (interpretation, feedback, and mutuality) interact with each other to shape the communications between the human and the AI. Outlining these elements and stages not only provide a content and process account of communications between humans and AI systems equipped with ToM-like capability, but also surface research opportunities to enhance such human-AI communication process to explore, understand, and examine these elements in specific stages. In the next section, we summarize two empirical studies inspired by the MToM framework to examine these elements during the construction and the recognition stages.

\begin{figure*}[h]
  \centering
  \includegraphics[width=0.6\linewidth]{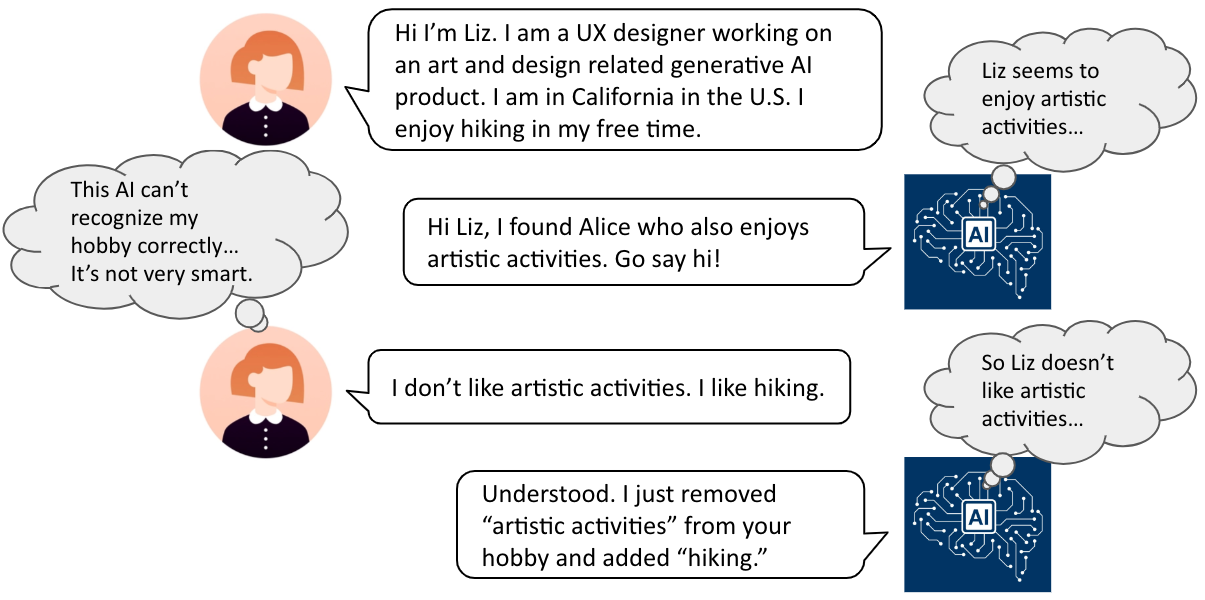}
  \caption{An example human-AI communication dialogue between a student Liz and an AI agent that can provide social recommendations based on Liz's self-introduction. This dialogue shows the recursive nature of the MToM in human-AI communication.} ~\label{fig:mtom-dialogue}
\end{figure*}

\section{Empirical Exploration of MToM in Human-AI Communication} \label{empirical_mtom}
Guided by the MToM framework, our work aims to enhance and understand the human-AI communication process by examining the interplay between interpretation, feedback and mutuality during the communication between humans and AI systems that exhibit ToM-like capability. Our work so far has focused on the recursive nature of ToM, specifically, second-level ToM, in human-AI communication, which is the idea of ``I can think about what you are thinking about me.'' So far we have examined second-level ToM in the first two stages of MToM: the construction stage and the recognition stage. At the construction stage, we examined the feasibility of how the AI can construct its interpretations of the user's interpretations of the AI; at the recognition stage, we explored the human's perceptions and reactions of the AI after recognizing the AI's interpretations of them, specifically when the AI's interpretations are wrong. We summarize our empirical work at these two stages in this section. 

Most of our work took place in online learning context, where AI agents are increasingly deployed as teaching assistants or social facilitators to provide informational and social support to online learners. Figure.~\ref{fig:mtom-dialogue} shows a human-AI communication dialogue between an AI agent acting as a social facilitator to provide social recommendations to online learners based on the inferences made about the student's social preferences from their self-introduction. 



\subsection{Constructing User's Perceptions of AI through Linguistic Cues}
Recent technical advancements have made it possible for AI systems to ``read users' minds'' by predicting our shopping preferences~\cite{linden2003amazon}, emotional states~\cite{schwartz2016predicting,wang2020sensing,de2013predicting}, and personalities~\cite{golbeck2011predicting} with fairly high accuracy. However, user's perceptions of such advanced AI systems have been under explored. Understanding user perceptions of such AI systems is critical to the success of human-AI communication, especially given people's often unrealistically high expectation of AI system's capability. This is especially common during the communications between humans and Conversational Agents (CAs), which can present human-level natural language understanding and generation when powered by LLMs, yet often present inconsistent performance across various task-specific capabilities. This ``gulf'' between user expectation and experience with CAs~\cite{Luger2016LikeAgents} has led to constant user frustration, frequent conversation breakdowns, and eventual abandonment of CAs~\cite{Luger2016LikeAgents,zamora2017m}. 

To understand how we could potentially mitigate or bridge this ``gulf'' between user expectation and experience with the CAs, we looked to the MToM framework for inspiration. We took inspirations from the recursive interpretations and feedback in the MToM framework, and envisioned an adaptive CA that could automatically construct a representation of the user's perceptions of the CA through verbal or behavioral cues embedded in user feedback. An automatic construction of the user's perceptions of the CA would enable the CA to monitor users' changing perceptions and adapt their behaviors accordingly to cater to users' needs, or even provide nudges to help users build a better mental model of the CA. 

To examine the feasibility of automatic construction of the user's perceptions of AI, we deployed an AI agent acting as a virtual teaching assistant to answer students' logistic questions about the class in an online class discussion forum for 10 weeks with about 376 students enrolled. We collected students' bi-weekly perceptions of the agent in terms of perceived anthropomorphism, intelligence, and likeability, as well as students' questions asked to the agent throughout. We then extracted various linguistic cues from students' questions asked to the agent, such as readability (the level of ease readers can comprehend a given text), sentiment (emotions conveyed through the language), linguistic diversity (diversity of the conversation topics or the richness of language used), and adaptability (how adaptable are students' questions to the agent's responses). We then built linear regression models using the linguistic characteristics as independent variables to predict each of the student community's perceptions of the AI agent (anthropomorphism, intelligence, likeability). 

We found that verbosity negatively associates with student perceptions of the AI agent, whereas readability, sentiment, diversity, and adaptability positively associate with anthropomorphism, intelligence, and likeability. Our findings suggest that it is feasible to extract linguistic features to measure users' perceptions of CA during conversations, and thus enable the CA to constantly interpret and provide desirable responses that cater to user perceptions. More details about this study and the model results can be found in our CHI 2021 paper~\cite{wang2021towards}.

\subsection{User Reactions and Perceptions of AI Misrepresentations}
Many hyper-personalized AI systems that can profile users' characteristics and traits have been deployed in people's daily lives, with the ultimate goal of providing personalized recommendations in shopping, music, social media, etc. As these systems become more advanced in profiling people's most personal and complex traits such as personalities and emotions~\cite{hall2017say,gou2014knowme,liao2021crystal}, they sometimes give the illusion of ``machines can read our minds''~\cite{guzman2020ontological}. This illusion has led to various---rather concerning---reactions and perceptions of AI with people attributing AI with beyond-human expertise at reading people's emotions and personalities~\cite{warshaw2015can,hollis2018being}. However, people's perceptions and reactions of AI when this illusion is broken in the face of~\textit{AI misrepresentations} have not yet been explored. 

AI misrepresentations refer to when AI systems make the wrong inferences about people's implicit characteristics and traits, such as personality, based on user data. Even algorithms with supposedly high accuracy can make mistakes when powering hyper-personalized AI systems in-the-wild~\cite{rao2015they}. Guided by the MToM framework, we situate this problem in the recognition stage of the MToM framework to examine people's reactions and perceptions of AI after recognizing that AI has an inaccurate interpretations of the human. Understanding people's reactions and perceptions of the AI after encountering AI misrepresentations could offer valuable insights into whether and how people changed their intuitions, beliefs, and reactions of AI in the face of AI fallibilities. This could provide critical implications for the future design and development of responsible interventions, mitigation, and repair strategies to retain user trust, minimize harms, and prevent overreliance when such AI systems inevitably err.

To understand people's reactions and perceptions of AI misrepresentations on their personalities, we conducted semi-structured interviews with 20 college students and a large survey experiment with 198 students on the Prolific platform. In both studies, we took a Wizard-of-Oz approach to fabricate intentionally inaccurate/accurate personality inferences based on participants' personality ground truth. We showed participants in both studies their ``AI-generated personality inferences'' to elicit their perceptions and reactions of AI misrepresentations. 

In both the interviews and survey experiment, we first familiarized participants with some sample student-AI dialogues where the AI agent provided a paragraph describing the student's personality based on a paragraph of the students' self-introduction. We measured participants' baseline perceptions of the agent after viewing the sample dialogues. In both studies, participants were randomly assigned to either receive accurate or inaccurate ``AI-generated personality inferences'' and their perceptions of the agent and reactions were measured after seeing their own inferences. We analyzed our interview data using reflexive thematic analysis, then built linear regression models and conducted moderation analysis to understand the changes in students' perceptions of AI before and after viewing AI misrepresentations. 

Our results showed that people's existing and newly acquired knowledge of AI are highly connected to people's reactions and perceptions of AI after encountering AI misrepresentations. Specifically, we found that participants acquired new knowledge from AI (mis)representations. Such newly acquired knowledge prompted participants to adopt different rationales to interpret how AI worked: AI works like a machine, human, and/or magic. These rationales could co-exist at any given time, yet are bounded by participants' existing AI knowledge, tech proficiency, and how much they could make sense of AI's specific inferences. Through our linear regression models, we also established that people's existing AI knowledge, i.e., AI literacy, can significantly moderate changes in people's trust of the AI after encountering AI misrepresentations, highlighting the importance of taking into account of people's knowledge and characteristics when building trustworthy AI systems~\cite{chen2023machine,chen2023understanding,ehsan2020human,schoeffer2022there}.

Based on our interviews, we found that people's AI knowledge, especially the rationales participants adopted after acquiring new knowledge from AI misrepresentations, are highly connected to participants' reactions to AI misrepresentations. After being shown the AI-generated inaccurate personality inferences about them, participants displayed a range of reactions: some participants believed there was some truth to AI misrepresentation; some participants rationalized it and blamed themselves instead; some participants were forgiving of the AI's mistakes. Building on top of prior work that has suggested people's tendency of over-trusting and viewing AI as an authority~\cite{kapania2022because,warshaw2015can,hollis2018being}, we highlighted that these reactions and perceptions still persisted, and even exacerbated, when people encountered AI misrepresentations. 

This work provides important implications of how AI systems can and should be designed to be aware of people's ever-evolving AI knowledge, and provide customized repair strategies accordingly to mitigate potential user harms from AI misrepresentations. We provided a specific set of rationales and encourage future work to explore techniques that could allow automatic identifications of users' rationales adopted in real-time. One mitigation strategy could be to provide explanations tailored to the specific rationale that people adopted at the time. For instance, if a user adopted the magic rationale, the AI could provide explanations to nudge the user to adopt the machine rationale to reduce overreliance.

\section{Discussion}


In the previous section, we presented two empirical studies inspired by the MToM framework to understand the construction and recognition stage of MToM in human-AI interaction. The first study shows that by leveraging the linguistic cues embedded in user feedback, it is feasible to equip the AI with a ToM to constantly model and interprets users' perceptions of the AI; The second study shows that by exploring users' reactions and perceptions of AI after recognizing AI's inaccurate ToM through AI feedback, we could provide more personalized and adaptive AI repair strategies to mitigate potential harms. Both studies contributed unique implications for enhancing human-AI communication by focusing on the interplay between the three elements (interpretations, feedback, mutuality) in the first two stages of the MToM framework. Here, we discuss other research opportunities that can be further explored to enhance human-AI interaction by examining the interplay between the three elements at each stage in the MToM framework. 

\textbf{Research Opportunities in the Construction Stage.}
At the construction stage, the AI system constructs its interpretations of the user's mind based on user feedback. In this stage of human-AI interaction, the following research questions can be asked by examining the three elements of user feedback, user feedback shaping AI interpretation, and AI interpretation, as seen in Figure~\ref{fig:mtom-framework}: (1) What kind of user feedback would help the AI construct accurate interpretations of the user's mind? (2) How can we construct AI's ToM through cues embedded in user feedback? (3) What should be constructed as part of the AI's interpretation of the user? Many existing research surrounding ToM in human-AI interaction have explored or begun to explore some of these questions. For example,~\citeauthor{baker2011bayesian} and many others have been exploring different techniques such as Bayesian ToM for the AI to detect and analyze observable or unobservable user feedback to model the human minds. Other fields such as emotion detection, ubiquitous computing, physical sensing can also offer valuable implications to the construction of AI's ToM. An underlying issue that should be examined as the foundation of these research questions is the problem of operationalization--- how can we operationalize ToM given the huge variations of human minds in different contexts across individuals? 

\textbf{Research Opportunities in the Recognition Stage.}
At the recognition stage, the user recognizes the AI's interpretations of the user's mind based on AI feedback. If we look at the three elements of AI feedback, AI feedback shaping user perceptions, and user perceptions at this stage (see Figure~\ref{fig:mtom-framework}), several research questions can be examined: (1) What kind of AI feedback can convey its interpretations of the user's mind to the user? (2) How can design features of the AI feedback shape or trigger user's perceptions if the AI? (3) What dimensions of user perceptions could change after recognizing AI's interpretations of their mind? A fundamental research question and opportunity here is to explore and map out the design characteristics/features of the AI feedback that could lead to changes in certain dimensions of the user perceptions of the AI. For example, certain wording or phrasing of the AI feedback could lead to increased levels of anthropomorphism~\cite{li2021machinelike}. Understanding and mapping out design features of AI feedback and changes in user perceptions could offer valuable implications to mitigate and even prevent harmful perceptions of AI. 

\textbf{Research Opportunities in the Revision Stage.}
At the revision stage, the AI system conducts revision of its interpretations of the user's mind based on user feedback, and then communicates the revision back to the user to re-establish mutual understanding in human-AI communication. The following research questions can be explored at this stage by looking at the three elements of user feedback, user feedback shaping AI's interpretations of the user's mind, and AI feedback shaping user's perceptions of the AI (see Figure~\ref{fig:mtom-framework}): (1) What kind of user feedback can help the AI revise its interpretations of the user's mind? (2) How can the AI take into account of user feedback to revise its interpretations of the user's mind? (3) What kind of AI feedback can help the user understand its revised interpretations of the user's mind? Existing work on human-centered explainable AI~\cite{ehsan2020human,ehsan2021expanding,liao2020questioning} have been trying to tackle question (3) by taking into account of user traits and characteristics when designing AI feedback that can explain AI's working mechanism to the user. Research efforts that aim at addressing question (1) and (2) are also underway, with some scholars seeking to mimic the human capability of metacognition and introspection~\cite{kawato2021internal, cox2005metacognition} when implementing AI systems. 

\section{Conclusion}
This paper proposed the Mutual Theory of Mind (MToM) framework to understand and design communications between humans and AI systems that are performing increasingly diverse social functions in human society. The MToM framework highlighted three key elements ( interpretations, feedback, mutuality) that continuously interact with each other throughout the three stages of the communication process (construction, recognition, and revision). The MToM framework thus provides a process and content account of human-AI communication as a iterative, mutual-shaping process of the human's and AI's interpretations of each other based on communication feedback. We summarized two empirical studies that were inspired by the MToM framework: the first study demonstrated an innovative use of user feedback to enhance AI's interpretation of user's perception of the AI during communication; the second study examined user perceptions and reactions after recognizing AI's incorrect interpretations of the user, highlighting the role of people's newly acquired and evolving knowledge in shaping human-AI communication. We concluded by discussing research opportunities in human-AI interaction through the lens of MToM at the construction, recognition, and revision stage by examining the interplay between the three key elements in the MToM framework. 

\begin{acks}
This research has been supported by US NSF Grant \#2247790 to the National AI Institute for Adult Learning and Online Education (AI-ALOE; \url{aialoe.org}).
\end{acks}

\bibliographystyle{ACM-Reference-Format}
\bibliography{reference}

\end{document}